# GENERALIZED MODIFIED PRINCIPAL COMPONENTS ANALYSIS OF RUSSIAN UNIVERSITIES COMPETITIVENESS

**Pavel Vashchenko – Alexei Verenikin – Anna Verenikina**

**Abstract**

The article is devoted to the competitiveness analysis of Russian institutions of higher education in international and local markets. The methodology of research is based on generalized modified principal component analysis. Principal components analysis has proven its efficiency in business performance assessment. We apply a modification of this methodology to construction of an aggregate index of university performance. The whole set of principal components with weighting coefficients equal to the proportions of the corresponding explained variance are utilized as an aggregate measure of various aspects of higher education. This methodology allows to reveal the factors which exert positive or negative influence on university competitiveness. We construct a kind of objective ranking of universities in order to estimate the current situation and prospects of higher education in Russia. It is applicable for evaluation of public policy in higher education, which, by inertia, aims to promote competition rather than cooperation among universities.

## Introduction and literature review

Educational service is a socio-economic category, a specific form of activity, which is based on providing the consumer with a set of services: information, knowledge, skills and competencies used for self-determination and personal development. Improving the quality of educational services of higher education institutions of the Russian Federation and increasing their competitiveness is one of the most important tasks in the process of modernization of the national education system.

In a knowledge-based economy, higher education becomes the most important factor of national competitiveness. It is there that the most productive human capital is formed, where knowledge and innovation are produced. Therefore, recent decades have been characterized by increasing global competition in higher education.

There is no doubt that there is a need for systemic indicators to assess the results of policies in the field of higher education, however, according to UNESCO report (Martin et al., 2011), there is currently no consensus on what 'quality' means in the higher education system as a whole due to the complexity, multi-purpose and multifunctional nature of higher education systems. Abovementioned UNESCO report describes approaches to developing systems of performance indicators for higher education systems. Such indicators determine different directions of higher education development. For example, the key activities are education, research and social development. Financial and infrastructural resources are important.

It's hard not to agree with Altbach, that university can be competitive in the case when it is provided with opportunities for engaging talented researchers, lecturers and students, with sufficient quantity and quality of material resources, infrastructural base, and with effective management model (Altbach, 2011). Global competitive advantages are achieved by universities in those countries where combined national strategies for competitive universities development are successfully implemented, and the level of expenditures on higher education per student is relatively high. The national system of higher education in Russia has great potential for development, a more productive utilization of which with appropriate government support can become a foundation for establishing globally competitive universities.

The main feature of the current state assessment of universities competitiveness is the fact that the efficiency is considered as the only indicator of the university competitiveness. University ranking and competitiveness are evaluated in different publications from Russia and abroad, we paid attention to the following researches. Shypulina, Gryshchenko and Bilenko (Shypulina et al, 2016) analyzed classification of international and national ratings depending on different classifications and evaluation criteria. Platonova, Fedotova, Musarskiy, Ulitina, Igumnov and Bogomolova (Platonova et al., 2016) identified the fact that Russian universities are focused on the performance indicators which use state bodies, which differ from the international assessment of competitiveness indicators. So, according to the authors, the system of intra-university evaluation should be reorganized. Bogoviz, Lobova, Ragulina and Alekseev (Bogoviz et al., 2019) criticized, that in Russia the same criteria for evaluating competitiveness of universities are used, regardless traditional or remote education form are provided by the university. Maslennikov, Grishina, Lyandau and Kalinina (Maslennikov et al., 2017) made empirical research for 42 Russian Universities from three groups: acting mainly on global, national or regional scale, and analyzed the strategies of one representative from each group.

The methodology of our research is based on principal component analysis (PCA), which is widely used in multidimensional statistics including university's competitiveness issues. For example, we can mention the research of Antipova, Shestakova and Melnik evaluated the performance indicators of universities and used PCA to choose the main factors for rating and performance auditing (Antipova et al., 2016). In the study made by Fuller, Beynon, and Pickernell (Fuller et al., 2019) on the basis of PCA investigated indicators of UK universities' Third Stream Activity, illustrating differences between entrepreneurial and enterprising university concepts. Bileviciute, Draksas, Nevera and Vainiute used descriptive statistical methods to analyze the use of modern innovation management methods that allowed university to survive in the difficult conditions of competition growth and state funding reduction (Bileviciute et al., 2019). Kosar and Scott critically examine university's Carnegie Classification (developed by the Carnegie Foundation for the Advancement of Teaching), its aggregate and per capita (per faculty member) variables and its use of two separate principal component analyses on each (Kosar, Scott, 2018).

## 1   Methodology, data and analysis

Competitiveness is a multidimensional characteristic that comprises a variety of indicators. Each, $i$-th indicator ($i = 1,…,n$; $n=34$ in this case) characterizes the performance of a $j$-th university ($j = 1,…,m$; $m=641$ in this case). Overall we deal with a matrix of initial data $X = \begin{pmatrix} x_{11} & \cdots & x_{1m} \\ \vdots & \ddots & \vdots \\ x_{n1} & \cdots & x_{nm} \end{pmatrix}$.

The key issue is how to choose appropriate weighting coefficients for the particular university activities $x_i$ that will not rely on subjective judgments.

We use principal component approach - a multidimensional statistical technique allows to put together diverse, almost incomparable factors. It transforms a set of original variables into a set of artificial uncorrelated variables: $Z = \begin{pmatrix} Z_1 \\ \vdots \\ Z_n \end{pmatrix} = \begin{pmatrix} z_{11} & \cdots & z_{1m} \\ \vdots & \ddots & \vdots \\ z_{n1} & \cdots & z_{nm} \end{pmatrix} = LX$, where $Z_1,…,Z_m$ are the first to $m$-th principal component vectors, $L = \begin{pmatrix} l_{11} & \cdots & l_{1n} \\ \vdots & \ddots & \vdots \\ l_{n1} & \cdots & l_{nn} \end{pmatrix}$ is the matrix of linear orthogonal transformation.

Principal component loadings are eigenvectors of the covariance matrix of initial data $\Sigma$: $(\Sigma - \lambda I)l_1^T = 0$. The corresponding characteristic equation $|\Sigma - \lambda I| = 0$ has $n$ real-valued nonnegative roots $\lambda_1 \geq \lambda_2 \geq … \geq \lambda_n \geq 0$ (eigenvalues of the covariance matrix $\Sigma$). The first principal component loadings are determined as the eigenvector that corresponds to the largest eigenvalue $\lambda_1$. The following principal components $Z_k = (z_{k1},…,z_{km})$ use as component loadings other eigenvectors that correspond to successively smaller eigenvalues $\lambda_k$, $k=2,…,n$. $\lambda_k$ is equal to variance of the $k$-th principal component. Total variance of principal components coincides with total variance of primary data, thus $\rho_k = \lambda_k / \sum_{k=1}^{n} \lambda_k$ is the share of total primary data variance explained by the $k$-th principal component.

The first principal component score $z_{1j}$ is known to be used as an aggregate indicator of activity of the $j$-th university. Unfortunately it explains only $\rho_1$ share of the variance of initial data and thus yields a substantive loss in exposing capability.

We use the generalized principal component approach approved by our previous research (Verenikin and Verenikina, 2019) to calculate an aggregate measure of regional environmental impact as a weighted sum of all principal component scores: $I_j = \sum_{k=1}^{n} \rho_k y_{kj} = \sum_{k=1}^{n} \rho_k \sum_{i=1}^{n} l_{ki}^2 x_{ij}$.

Note that we use here modified principal component scores $y_{kj} = \sum_{i=1}^{n} l_{ki}^2 x_{ij}$ instead of ordinary principal components $z_{kj}$ (Aivazian, Stepanov, Kozlova, 2006). This makes it possible to avoid negative principal component scores as constituting elements of the composite index. The modified principal components $y_{kj}$ are weighted by the corresponding shares of explained variance $\rho_k$. There is no loss in variance of the considered data. The explaining capability of the proposed indicator is extended to the total variance of initial variables. The distinguishing feature of the proposed composite measure is that it is not sensitive to subjective preferences concerning the relative significance of specific factors of university competitiveness.

The data are normalized within the range from one to ten in order to obtain the uniform increasing impact of all the factors of concern on the level of the resulting aggregate index. We adjust it to 1-10 ranking scale in the following way: $x_{ij}^n = 1 + 9\left(\frac{x_{ij} - x_{ij}^{\min}}{x_{ij}^{\max} - x_{ij}^{\min}}\right)$, where $x_{ij}^n$ is a normalized variable, $x_{ij}^{\max}$ and $x_{ij}^{\min}$ are correspondingly the "best" and the "worst" value of initial indicator $x_{ij}$.

The analysis is focused on data from open official statistics, mainly from official web site of Department of State Policy in Higher Education and Youth Policy of the Russian Ministry of Science and Higher Education, dedicated to Monitoring of efficiency of higher education institutions of Russia (http://indicators.miccedu.ru/monitoring) for the year 2017.

We excluded from the sample those universities that did not accept students for Bachelor's degree programs, as well as those universities that had been deprived of their licenses for educational activities at the time of preparing this article.

Original data is grouped into a number of subsets or pillars that reflect definite attributes of university competitiveness. The rating considers a number of indicators that reflect different factors of university activity. They were grouped into five pillars called: «Student body», «Research» (R&D activity), «International activities», «Academic funds» (finance and infrastructure) and «Academic staff» (see Tab. 1).

**Tab. 1. Indicators and pillars**

| Pillar | Indicator |
|---|---|
| **A. Student body** | A1. Average minimum Unified State Exam score of students enrolled in bachelor and specialist programs |
| | A2. Number of Olympiad winners admitted to bachelor and specialist programs |
| | A3. Share of students enrolled in internship, master and PhD programs |
| | A4. Share of students enrolled in internship, master and PhD programs with diplomas of other institutions |
| | A5. Number of postgraduate students and interns (per 100 students) |
| | A6. Share of listeners from the external organizations in total number of the listeners under programs of professional training and upgrading |
| **B. Research** | B1. Number of citations of publications published over the past 5 years, indexed in the Web of Science (per 100 scientific and pedagogical workers) |
| | B2. Number of citations of the publications published for last 5 years, indexed in Scopus (per 100 academic workers) |
| | B3. Number of publications, indexed in Web of Science (per 100 academic workers) |
| | B4. Number of publications, indexed in Scopus (per 100 academic workers) |
| | B5. Total volume of R&D, thousand rubles |
| | B6. Share of income from R&D in total income of educational organization, % |
| | B7. Share of revenues from R&D performed in-house (without co-authors) |
| | B8. Revenues from R&D, excluding budgetary funds (per 1 academic worker) |
| | B9. Share of academic staff up to 30 years old, Candidates of Science - up to 35 years old, Doctor of Science - up to 40 years old in the total number of academic staff |
| | B10. Number of scientific journals, published by the educational organization |
| | B11. Number of grants received during the reporting year per 100 academic workers |
| **C. International activities** | C1. Share of foreign students (except for the CIS countries) enrolled in bachelor, specialist, master programs |
| | C2. Share of foreign students who have completed bachelor, specialist and master programs |
| | C3. Share of bachelor, specialist and master programs students, who have studied abroad for at least 1 semester |
| | C4. Number of bachelor, specialist, master programs students of foreign educational organizations who have studied in the educational organization, at least one semester |
| | C5. Number of leading foreign professors, teachers and researchers working in the educational organization at least 1 semester |
| | C6. Share of foreign citizens (except for CIS countries) |
| | C7. Amount of funds received from R&D performed by foreign citizens and legal entities |
| | C8. Amount of funds from educational activities received from foreign citizens and legal entities |
| **D. Academic funds** | D1. Revenues of the educational organization from income-generating activities per one academic worker |
| | D2. Share of incomes from profitable activities |
| | D3. Ratio of average academic salary to average regional salary |
| | D4. Incomes per student |
| | D5. Number of personal computers per student |
| | D6. Number of copies of printed publications in the libraries per student |
| **E. Academic staff** | E1. Share of PhDs holders in the total academic staff |
| | E2. Number of Candidate and Doctor of Science degree holders per 100 students |
| | E3. Share of full-time academic workers in the total number of academic staff |

Source: composed by the authors.

## 2 Results and discussion

The rating is headed by Russian largest, systemically important and internationally recognized Universities. Among top-15 there are 7 universities representing Central Federal District of Russia (Moscow city), 4 universities – North-Western District (Saint Petersburg city), 3 universities – Siberian (Tomsk and Novosibirsk cities), and one – Volga Federal District (Kazan city). Only one university in top is in private ownership (NES). Among the outsiders of our rating are mostly theatrical universities and private regional universities (see Tab. 2).

It is worth mentioning that the development of university education goes hand in hand with economic agglomeration and urbanization. In particular, the history and culture of the Russian capital cannot be understood without considering the role of MSU, SPbU and National Research Universities. But the development of the country's largest universities cannot be imagined in isolation from its home city. The progress of modern megacities and territorial and industrial complexes is largely based on the effect of scale, which, in particular, are subject to investment in infrastructure - transport and communications, industrial and technological, scientific and technical and socio-cultural.

At the macro level, a characteristic feature of the knowledge economy is complementarity and synergy between academic science and education. MSU, SPbU and National Research Universities are initiators and active participants in these integration processes. The effect of the scale of systemically important branches and sectors of the knowledge economy, in particular, the education system, also objectively determines the current processes of integration of regional universities, which results in a system of large universities of federal significance.

Taking into account the general regularities of economic development, in which the effects of scale and diversity play an important role, one can expect in the future an organic combination of processes of specialization and diversification of activities in the sphere of university education and science. The established large, authoritative scientific schools can become the basis here. Their development, inevitably accompanied by the emergence of new research and training centers - "growth points" of the new economy, should become an important competitive advantage of individual universities and the Russian higher education system on an international scale.

Therefore, the task of identifying and targeting support for fundamental, promising areas of scientific and pedagogical activity is becoming a priority.

Competitiveness index is a linear combination of the whole set of modified principal component scores: $I_j = \sum_{k=1}^{n} \left( \lambda_k \sum_{i=1}^{n} l_{ki}^2 x_{ij} \right) / \sum_{k=1}^{n} \lambda_k$. So it can be considered as a composition of partial indices which sum up weighted modified principal component scores for each data pillar. These sub-indices generate the university rankings with respect to particular pillars (see Tab. 3). They provide a glimpse of the factors of university competitiveness and of the potential for its improvement.

**Tab. 2. The overall ranking of Russian universities: leaders and outsiders**

| Rank | Leaders | Rank | Outsiders |
|---|---|---|---|
| 1 | National Research Nuclear University – «MEPhI» | 627 | Smolensk Orthodox Theological Seminary |
| 2 | M.V.Lomonosov Moscow State University – «MSU» | 628 | Northern Institute of Entrepreneurship (private) |
| 3 | St. Petersburg University – «SPbU» | 629 | Dagestan Medical Dental Institute (private) |
| 4 | National Research Tomsk Polytechnic University – «TPU» | 630 | Volgograd State Institute of Arts and Culture |
| 5 | The Moscow Institute of Physics and Technology – «MIPT» | 631 | M.P. Mussorgsky Urals State Conservatory |
| 6 | St. Petersburg National Research University of Information Technologies, Mechanics and Optics – «ITMO University» | 632 | Ekaterinburg State Theatre Institute |
| 7 | The New Economic School – «NES» | 633 | Institute of Television, Business and Design |
| 8 | National Research University Higher School of Economics – «HSE University» | 634 | Siberian Independent Institute (private) |
| 9 | Saint Petersburg National Research Academic University of the Russian Academy of Sciences – «Alferov University» | 635 | Kuban Medical Institute (private) |
| 10 | National Research Tomsk State University – «TSU» | 636 | Smolensk State Institute of Arts (regional) |
| 11 | National University of Science and Technology – «MISIS» | 637 | Novosibirsk State Theatre Institute |
| 12 | Peoples' Friendship University of Russia – «RUDN University» | 638 | Institute of Economics and Law (Nazran) – (private) |
| 13 | Novosibirsk State University – «NSU» | 639 | Institute of Theatrical Arts, Moscow |
| 14 | Peter the Great St. Petersburg Polytechnic University – «POLITECH» | 640 | Yaroslavl State Theatre Institute |
| 15 | Kazan Federal University – «KFU» | 641 | Volgograd Conservatory named after P.A. Serebryakov |

Source: composed by the authors

**Tab. 3. Pillars: leaders**

| Rank | Pillar A. Student body | Pillar B. Research | Pillar C. International activities | Pillar D. Academic funds | Pillar E. Academic staff |
|---|---|---|---|---|---|
| 1 | St. Petersburg University – «SPbU» | National Research Nuclear University – «MEPhI» | RUDN University | The New Economic School – «NES» | Russian Academy of Entrepreneurship |
| 2 | National Research University Higher School of Economics – «HSE University» | The Moscow Institute of Physics and Technology – «MIPT» | National Research Nuclear University «MEPhI» | Jewish university, Moscow | M.V.Lomonosov Moscow State University – «MSU» |
| 3 | Saint Petersburg National Research Academic University of the Russian Academy of Sciences – «Alferov University» | Novosibirsk State University – «NSU» | National Research Tomsk Polytechnic University – «TPU» | Eastern Academy of Economics, Law and Humanities - «Academy VEGU» (private) | Russian Academy of Advocacy and Notaries (private) |
| 4 | Diplomatic Academy of the Ministry of Foreign Affairs of the Russian Federation | ITMO University | St. Petersburg University – «SPbU» | Humanitarian Institute | National Business Institute |

| 5 | Moscow Higher School of Social and Economic Sciences | National University of Science and Technology – «MISIS» | Kursk State Medical University | The National Open Institute in St. Petersburg | St. Petersburg University – «SPbU» |
|---|---|---|---|---|---|
| 6 | ITMO University | M.V.Lomonosov Moscow State University – «MSU» | Glinka Nizhny Novgorod State Conservatoire | Humanitarian Social Institute | A.I. Yevdokimov Moscow State University of Medicine and Dentistry |
| 7 | The New Economic School – «NES» | National research Tomsk State University – «TSU» | Moscow State Linguistic University | Institute of Social Sciences | Kazan State Academy of Veterinary Medicine named after N.E. Bauman |
| 8 | Moscow Humanitarian and Technical Academy (private) | National Research Tomsk Polytechnic University – «TPU» | Peter the Great St. Petersburg Polytechnic University - «POLITECH» | Russian Academy of Entrepreneurship | Institute of Social Sciences |
| 9 | Russian Orthodox University of Saint John the Divine | Saint Petersburg National Research Academic University of the Russian Academy of Sciences – «Alferov University» | Moscow State Institute of International Relations – «MGIMO University» | Moscow Innovation University (private) | Higher School of Folk Arts (Academy). |
| 10 | North-Western State Medical University named after I.I. Mechnikov | Stavropol State Agrarian University | Far Eastern Fisheries University – «FESTFU» | Moscow Psychological and Social University | International Institute for Economics and Humanities Machon Chamesh |

Source: composed by the authors

It turns out that the leaders of the rating have huge potential of improvement of their activities. They have a long way to go to gain high positions in almost all pillars. For instance, MSU, the oldest and most prominent Russian university, which is ranked second in overall rating – is present among the leaders in two pillars only: in academic staff it has got the second and in research – the sixth rank. It relays on its own faculty and has no intention to engage in international turnover of academic staff. This reflects huge reproduction of its scientific and teaching potential but hinders its renovation in the long run. Private universities are dominating the state-owned institutions in salaries of academic staff and in fundraising. It yields higher ratios of academic staff with doctor and doctor habil. degrees. Still the state-owned universities provide public financing for their students and thus gain the leading positions in this pillar.

## Conclusion

In Russia the state system of the assessment of competitiveness of Universities are still under construction. Since 2012 authorities are used the indicators, threshold values and criteria on the basis of which the selection of the Universities, having signs of inefficiency is carried out by the Ministry of Education and Science. We used these indicators and generalized modified PCA to construct the rating of university competitiveness.

Unlike most of the existing university competitiveness ratings, our rating methodology reflects a comprehensive approach to assessing competitiveness. The rating provides an integral assessment of the current state of Russian universities competitiveness. We don't use any expert assessments and impose any subjective weights to the factors.

The results of this research are potentially useful for both scholars and practitioners. In fact, our research lays the foundation for regular (e.g., once in several years) consideration of Russian universities competitiveness. Investigation of the factors that determine university ranking can be used to improve competitiveness of Russian universities.

As a matter of further research, the inclusion of new indicators in our index, can serve to improve the comprehensive index of universities competitiveness.

## Acknowledgments

The study was supported by RFBR, research project No.18-010-00974A "Developing the model of management of territorial resource potential".

The publication was prepared with the support of the «RUDN University Program 5-100».

**Contacts**

Pavel Vashchenko

M.V. Lomonosov Moscow State University

Leninskie gory, 3rd new educational building, Moscow, 119991, Russian Federation

pavelvashchenko@outlook.com

Alexei Verenikin

M.V. Lomonosov Moscow State University

Leninskie gory, 3rd new educational building, Moscow, 119991, Russian Federation

verenikin@mail.ru

Anna Verenikina

Peoples' Friendship University of Russia (RUDN University)

6 Miklukho-Maklaya Str., Moscow, 117198, Russian Federation

verenikina_ayu@pfur.ru